# HOW TO MEASURE AND MODEL LIGHT-INDUCED SPIN TRANSFER

Sinéad A. Ryan[1]†*, Mohamed F. Elhanoty[2]†, Anya Grafov[1], Peter C. Johnsen[1], Na Li[1], Justin M. Shaw[3], Anna Delin[4,5,6], Anastasios Markou[7,8], Edouard Lesne[7], Claudia Felser[7], Olle Eriksson[2,9], Erna K. Delczeg-Czirjak[2,9], Debjani Karmakar[2,10,11], Henry C. Kapteyn[1,12], Oscar Grånäs[2], and Margaret M. Murnane[1]

†Equal contributors
*Corresponding author: sinead.ryan@colorado.edu

[1]JILA, University of Colorado Boulder, 440 UCB, Boulder, CO 80309, USA

[2]Division of Materials Theory, Department of Physics and Astronomy, Uppsala University, Box-516, SE 75120, Sweden

[3]Applied Physics Division, National Institute of Standards and Technology, Boulder, CO 80305, USA

[4]Department of Applied Physics, School of Engineering Sciences, KTH Royal Institute of Technology, AlbaNova University Center, SE-10691 Stockholm, Sweden

[5]Swedish e-Science Research Center (SeRC), KTH Royal Institute of Technology, SE-10044 Stockholm, Sweden

[6]Wallenberg Initiative Materials Science for Sustainability (WISE), KTH Royal Institute of Technology, SE-10044 Stockholm, Sweden

[7]Max Planck Institute for Chemical Physics of Solids, 01187 Dresden, Germany

[8]Physics Department, University of Ioannina, 45110 Ioannina, Greece

[9]Wallenberg Initiative Materials Science for Sustainability (WISE), Uppsala University, 75121 Uppsala, Sweden

[10]Technical Physics Division, Bhabha Atomic Research Centre, Mumbai 400085, India

[11]Homi Bhabha National Institute, Trombay, Mumbai 400085, India

[12]KMLabs Inc., Boulder, CO, USA

ABSTRACT

Femtosecond laser light can transfer spin angular momentum between magnetic subspecies that exhibit hybridized valence bands within an alloy or compound, and represents the fastest route for manipulating the magnetization of a material. To date, ultrafast spin transfer has predominantly been explained in terms of the initial and final states available for laser excitation. Here, by comparing the measured and calculated dynamics across the entire $M$-edges of two very similar Heusler compounds, $Co_2MnGa$ and $Co_2MnGe$ as well as a sample of elemental Co, we find that simply accounting for the initial and final electron states available for laser excitation cannot alone explain the experimental observations. The influence of spin lifetimes must also be included, due to the shifting of the Fermi level upon replacing Ga with Ge, or the presence of crystalline disorder. This explains why the ordered $L2_1$ phase of $Co_2MnGa$ demonstrates strong laser-induced magnetic signal enhancements across the entire Co-edge, while similar enhancements were not observed in partially disordered $Co_2MnGe$. Although intra-site spin-transfers were expected in the minority channel in pure Co due to the presence of many more available states in the minority channel above the Fermi level, no such signal was observed due to very short few-femtosecond spin lifetimes in a metal. Finally, we identify key regions in the magnetic asymmetry where a transiently enhanced signal could be misinterpreted as a light-induced spin-transfer signature.

INTRODUCTION

As the global economy becomes increasingly dependent on big data and green energy, there is an ever-growing need for high-quality magnetic materials for data storage(*1*), wind power applications, and in electric vehicles(*2*). Approximately 90% of cloud storage utilizes perpendicular magnetic recording(*3*) and our data storage requirements in 2025 may be more than 90 times greater than in 2010(*4*). One study noted that meeting global wind power targets for 2050 will require an 11- to 26-fold increase in world production of Rare Earth elements, to facilitate the transformation of mechanical energy to electricity(*2*). One promising avenue for future magnetic technologies is the implementation of spintronics: by combining electronic and magnetic properties of electrons in technological applications, more efficient data storage architectures can be created(*5*). Ultrafast spintronics, where magnetic states are controlled via very short laser pulses, could facilitate control on much faster timescales than any existing technologies(*6*). However, such advances will require a complete understanding of ultrafast magnetism from Ångström length scales to attosecond time scales.

It has been almost 30 years since Beaurepaire *et al.*(*7*) uncovered subpicosecond demagnetization of Ni. However, many open questions still remain about how light can be used to manipulate magnetic materials. These questions include: how is angular momentum dissipated on such short timescales(*8–11*), what are the dominant demagnetization pathways(*12–16*), and how does one correctly measure, model, and interpret the signatures of ultrafast magnetism(*17–20*)?

A recent exciting topic is light (optically) induced spin transfer (or OISTR)(*21*), in which femtosecond laser light can excite spin transfer between different elements that exhibit hybridized valence bands within an alloy or compound. By probing the resultant dynamics using broadband extreme UV (EUV) high harmonic combs, the element-specific ultrafast magnetic response can be extracted(*22, 23*). These light-induced spin transfers are technologically promising as they provide a mechanism to enhance the magnetization of an element on a timescale of a few-femtoseconds or less – much faster than traditional magnetization switching techniques. However, recent papers(*24–26*) have called into question some of the interpretations found in OISTR experiments(*22, 23*), where transient enhancements in the magneto-optical signal were directly attributed to OISTR. This motivates the need for more complete methods to both measure and model light-induced spin transfer.

To date, most theoretical models for OISTR were based on calculations of the initial and final electron states available for laser excitation. Moreover, all but one experiment(*24*) investigating OISTR effects measured the magnetic asymmetry of the relevant magnetic sublattices at only one or two EUV probe energies, i.e. they sampled only a small fraction of the resonant *M*-edges by measuring one narrow energy region for each element. $Co_2MnGe$ was one of the first materials used to investigate the OISTR effect with element-specificity by Tengdin *et al.*(*22*) using single probe energies at the Co and Mn edges. More recently, a study of spin transfer in $Co_2MnGa$ identified strong energy-dependent OISTR signatures in the magneto-optic Kerr effect (MOKE) signal across the entire *M*-edges of the two magnetic sublattices involved (Ryan *et al.*(*24*)). However, surprisingly, extensive measurements of elemental Ni, where OISTR is not possible (because only one element is present), at various probing energies and with different geometries, appeared to observe similar magnetic enhancement signatures(*19, 25, 27*). Furthermore, enhancements were measured in an FeNi sample even when the geometry was chosen to prevent direct optical excitation of the material(*26*). These experiments conclusively demonstrate that a magneto-optical enhancement at a single EUV probe energy is not sufficient evidence to demonstrate OISTR. These experiments also raise fundamental questions about the interpretation of apparent signatures of magnetic enhancement at different EUV probe energies.

In recent work, we developed an approach to both measure and calculate the induced spin dynamics across the entire *M*-edges of coupled sublattices, in this case, Co and Mn in $Co_2MnGa$ (Ryan *et al.*(*24*)). This allows for very detailed comparisons between theory and experiment, to distinguish the dominant processes that occur during and after excitation of the material by light, e.g. the balance between spin transfer, spin scattering, and demagnetization, and how these processes depend on the laser fluence.

In the present work, we compare the measured and calculated energy-dependent dynamics across the entire *M*-edges of two seemingly very similar half-metallic Heusler compounds, $Co_2MnGa$ and $Co_2MnGe$. We observe that simply accounting for the initial and final electron states available for laser excitation cannot explain the experimental observations or provide a comprehensive understanding. Surprisingly, while the ordered $L2_1$ phase of $Co_2MnGa$ demonstrates strong laser-induced magnetic signal enhancements across the entire Co-edge, similar enhancements were not observed in the partially disordered B2 phase of $Co_2MnGe$. We attribute the weaker response to reduced spin lifetimes that arise from a combination of the shifting of the Fermi level upon replacing Ga with Ge leading to stronger spin-orbit coupling, as well as the presence of disorder. To make a more comprehensive comparison, we also measure elemental Co to compare the MOKE signatures from a material where intersite spin transfers are not possible; however, same-site (intrasite) optical excitations are predicted to be strong because there are many more available states in the minority channel above the Fermi level compared to the majority channel. However, no intrasite spin-transfer signal is observed in Co, due to the very short ~5-20-femtosecond electron-spin lifetimes in 3d ferromagnets such as Co and Ni(*28–30*). Our large data set, recorded for three different materials, allows us to conclude that engineering of the valence bands through elemental substitution and alloying can be used to tune the ultrafast response of materials to light, enabling a deeper understanding of when intersite transfers (i.e. OISTRs), same-site spin transfers, or other magneto-optical effects are dominant, or how their contributions change with excitation fluence. Finally, we identify key regions in the dynamic magnetic asymmetry where a transiently enhanced signal could be misinterpreted as a light-induced-spin-transfer-like signature.

## RESULTS

Our choice for comparing these two Heusler compounds is motivated by four reasons. First, Heusler materials show a very strong Slater-Pauling relationship(*31*). Therefore, the addition of an extra valence electron in the unit cell of $Co_2MnGe$, compared to $Co_2MnGa$, results in a known magnetic moment of 4.94 $\mu_B$ per unit cell, which is significantly larger than the measured value of 4.06 $\mu_B$ for $Co_2MnGa$(*31*). Second, Ga and Ge are neighboring elements on the Periodic Table, and the density of states (DOS) is half-metallic and rather similar for the two, except for the difference in band filling. Both systems have electron states near the Fermi level that are relevant for optical excitation by the laser pump pulse, dominated by Co- and Mn-projected 3d states. Therefore, substituting Ge with Ga allows engineering of the band structure while preserving the main features of the electron states, and the half metallicity of the two Heusler compounds allows for a direct comparison between the two materials (see Figs. 1(a) and (b)). Third, substituting Ge with Ga leads to a red shift of ≈ 0.4 eV of the Fermi level, as shown in Figs. 1(a) and (b), which modifies the valence DOS and impacts the interpretation of the measurements. Fourth and finally, Heusler compounds have generated a great deal of recent research interest as they have excellent chemical stability, versatile tunable ground state properties, and the ability to host topological states as well as having promising applications in spintronics(*32*).

In our study, a 40 fs - 55 fs, ~800 nm pump pulse excites the sample, and the resultant dynamics are probed with EUV TMOKE at the *M*-edge. Equations for the EUV TMOKE asymmetry are given in the *Methods* section. The EUV probe is generated via high harmonic generation (HHG) and has a duration of <20 fs. More details of the experimental techniques are given in the *Methods* section. It is important to note that while the $Co_2MnGa$ sample was in the fully ordered $L2_1$ phase, the $Co_2MnGe$ was in the partially disordered B2 phase, and the Co sample was polycrystalline. Time-dependent density functional theory (TDDFT) calculations assumed perfect order for all systems (i.e. $L2_1$ phase for both Heuslers and HCP for Co). The effects of disorder on the experimental results are discussed below, and more details of the samples and their growth methods are given in the *Methods* section.

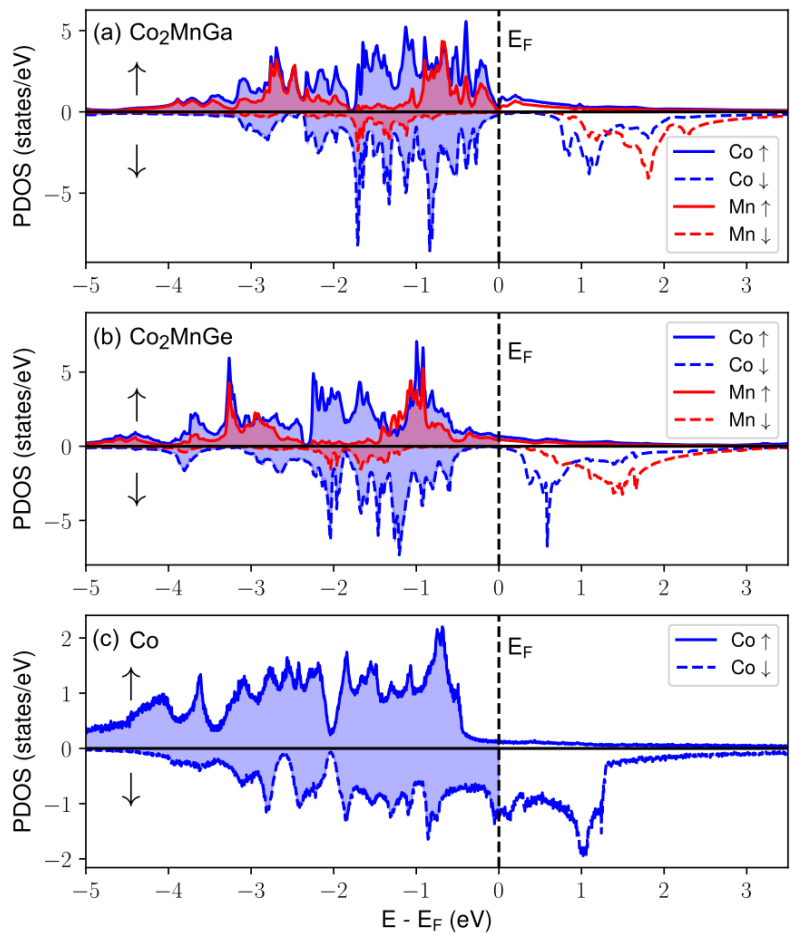

**Figure 1:** The calculated spin-resolved projected density of states (PDOS) of the 3d states separated by elemental contributions for $L2_1$ phases of (a) $Co_2MnGa$ (b) $Co_2MnGe$ and (c) HCP phase of pure Co. A pump excitation of approximately 1.55 eV excites electrons across the half-metallic gaps of $Co_2MnGa$ and $Co_2MnGe$ (see main text).

The density of states (DOS) in Fig. 1 shows that both $Co_2MnGa$ and $Co_2MnGe$ are half-metallic, i.e. they have a gap in the minority spin channel. The pure Co sample (Fig. 1(c)) is metallic in both spin channels and therefore has available states at the Fermi energy in both the spin-up and the spin-down channel. In all three materials ($Co_2MnGa$, $Co_2MnGe$, and pure Co), there is a large number of available states above the Fermi energy, primarily in the spin-down states. Therefore, we expect many excitations in the spin-down channel. Fig. 1 also demonstrates that unoccupied states that are closest to the Fermi level are associated with Co 3d orbitals. Furthermore, within the energy range of the electronic excitations that could be induced by the pump laser (1.54 eV - 1.60 eV photon energies), there are strong excitation pathways in both Heuslers where electrons from Co minority states below the Fermi energy could be excited into (hybridized) Mn minority states above the Fermi energy (see Fig. S6). This Co-Mn spin-transfer effect transiently increases the Co magnetization while reducing the Mn magnetization in both the Co$_2$MnGe and the Co$_2$MnGa samples(*22*, *24*) with distinctive signatures in each material, as our analysis presented below and the data of Fig. 2 demonstrate.

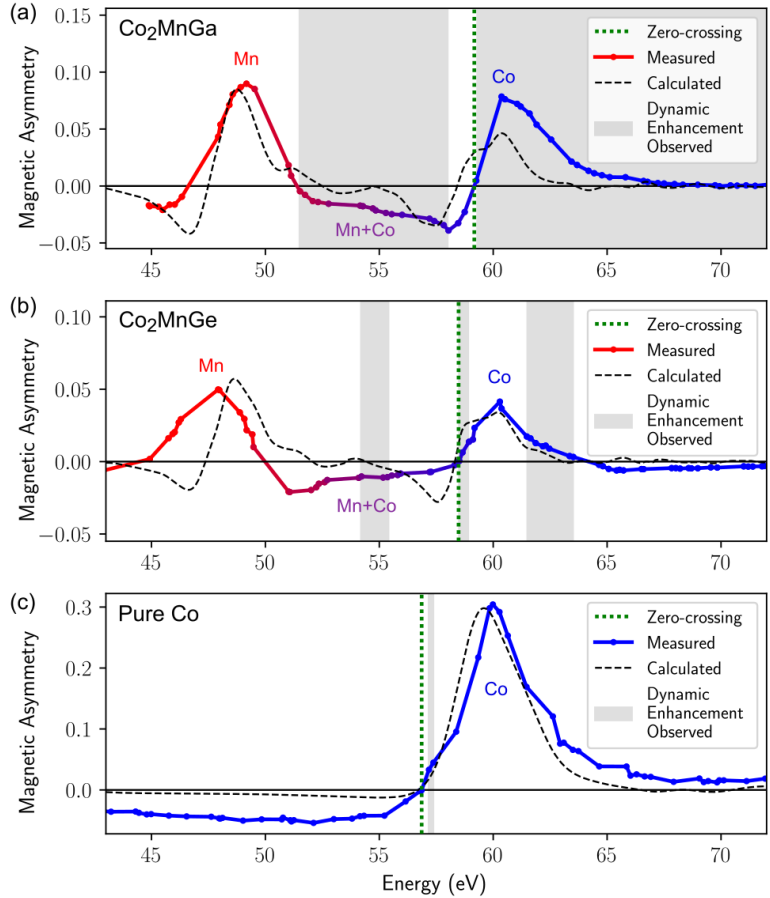

**Figure 2:** The measured and calculated ground state asymmetries. Gray bars indicate regions where the measured asymmetry exhibits a transient enhancement following laser pumping with an incident fluence of 3.4 mJ/cm². The corresponding absorbed fluences are 2.5 mJ/cm² for both Co and Co$_2$MnGe, and 2.2 mJ/cm² for Co$_2$MnGa. Zero-crossing regions which are addressed in the main text are marked with green dotted lines. (a) The measured and calculated ground state asymmetry for a Co$_2$MnGa L2$_1$ crystal. (b) The measured ground state for the B2 crystal phase of Co$_2$MnGe. The calculated asymmetry assumes no crystalline disorder (i.e. the L2$_1$ structure). (c) The measured asymmetry from the polycrystalline pure Co sample and the calculated asymmetry for an ordered HCP Co lattice. Amplitudes of the theoretical curves were scaled to match the measured data.

Fig. 2 shows the measured and calculated ground state asymmetries for the three materials. Fig. 2(a) shows good qualitative agreement between the theoretical and experimental asymmetry curves of Co$_2$MnGa(*24*). However, in Fig. 2(b), the shape of the measured ground state asymmetry of Co$_2$MnGe differs from what is predicted by DFT – especially in the central energy region from 51 eV – 57 eV. In the measured data, this energy region has a positive slope whereas the theoretical calculation of the asymmetry exhibits a negative slope. In this region between the Mn and Co M-edges, the polarization-dependent reflectivity ($I_{\pm}$) (see Eq. 1-3 in Methods) has mixed contributions from both Co and Mn. Identifying this region of overlap between the Co and Mn both in theory and experiment is very crucial because time-resolved measurements in this overlap region will not achieve element-specificity and may contain contributions both from Mn and Co. This sets our first constraint to correctly interpret the element-specific time-resolved magneto-optical data of alloys and compounds– by avoiding measuring the induced dynamics in the regions of known overlap between different elements.

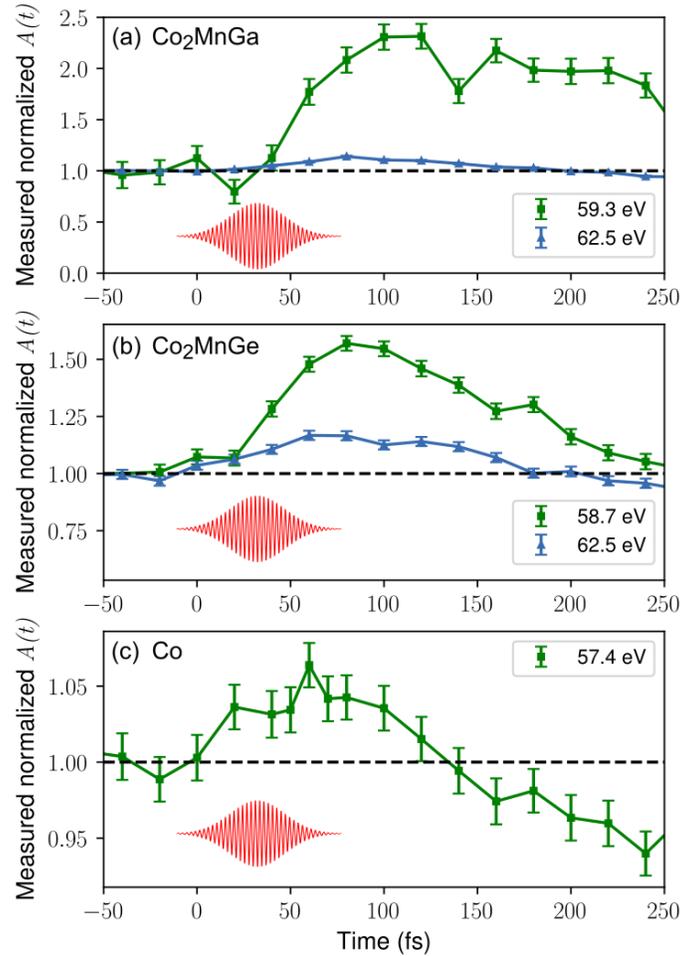

**Figure 3:** Measured transient MOKE signals which show enhancement for (a) $Co_2MnGa$, (b) $Co_2MnGe$ and (c) pure Co. The incident laser fluence was 3.4 mJ/cm². The corresponding absorbed fluences were 2.5 mJ/cm² for both Co and $Co_2MnGe$, and 2.2 mJ/cm² for $Co_2MnGa$. Green curves (square data points) represent enhancements seen on the positive side of the zero-crossing of each sample (regions marked with green vertical dashed lines in Fig. 2). The two Heusler samples have additional enhancement curves (blue triangles) which were measured at 62.5 eV, i.e. on the positive side of the Co peak.

In Fig. 2, every region where a transient enhancement in the MOKE signal was measured is highlighted in gray. Transient enhancements measured in the MOKE signal have previously been used as a signature of the OISTR effect(*22, 23*). However, enhancements have also been measured in pure elements (where OISTR is not possible) in specific energy regimes(*19, 25, 27*). Therefore, to disentangle the origins of these enhancement signatures, one needs to carefully compare the behavior of the measured samples with theory; here we used TDDFT calculations. Signals from regions of enhancement for each of the three materials investigated here are depicted in Fig. 3. Additional transient MOKE signals at other energies for pure Co and Co$_2$MnGe are given in the Supplementary Materials (SM), Figs. S1-S4, while data for Co$_2$MnGa appears in the SM of our previous work(*24*).

The original measurements of the B2 phase of Co$_2$MnGe(*22*) hypothesized a 10% spin-transfer based enhancement of the Co moment based on the transient MOKE enhancement measured at the Co-edge. We re-investigated this sample by tuning the photon energies of the EUV harmonics and the laser fluence, to probe the entire Co and Mn edge regions. We find substantial differences in the dynamic MOKE signal across different probing energies at the *M*-edge for the Co$_2$MnGe. While enhancements of 10% or more can be seen at specific probe energies in our new data (Figs. 2(b) and 3(b)), as well as in the original study(*22*)), surprisingly, at many energies, very small or even vanishing enhancements were measured, including at the Co peak itself. This is in contrast to Co$_2$MnGa (Fig. 2(a)), where enhancements in the magnetic signal appear across the entire *M*-edge of Co, although the magnitude of the enhancements also show strong variation.

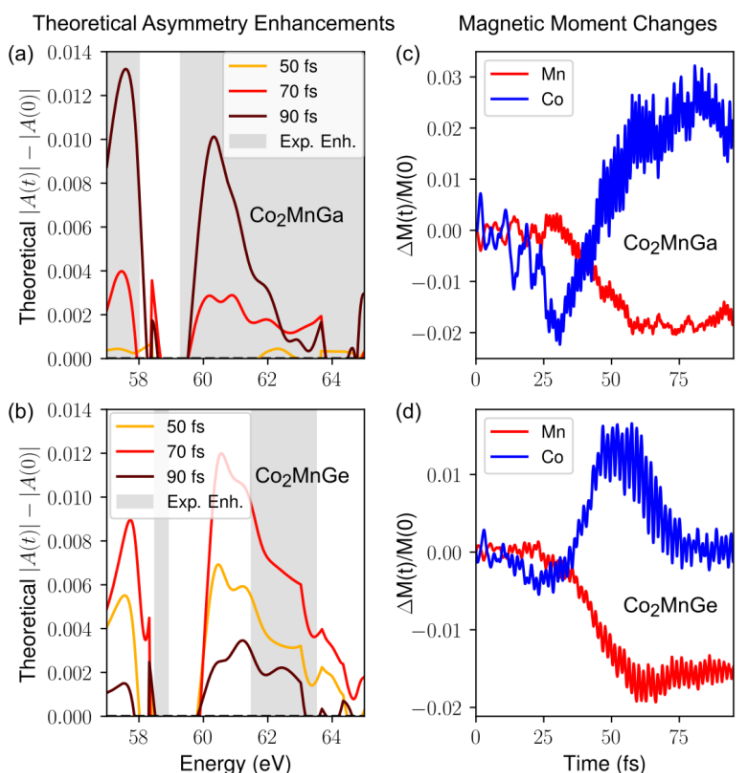

**Figure 4:** Theoretical asymmetry enhancements and changes in magnetic moment. The theoretically calculated magnitudes of positive asymmetry enhancements are given at the Co-edge for (a) Co$_2$MnGa and (b) Co$_2$MnGe. Grey bars show where enhancements were measured in the experimental data. The calculated changes in the magnetic moment with time are given on the RHS for (c) Co$_2$MnGa and (d) Co$_2$MnGe. Experimental absorbed fluences of 2.5 mJ/cm$^2$ for Co$_2$MnGe, and 2.2 mJ/cm$^2$ for Co$_2$MnGa were used for the calculations (based on 3.4 mJ/cm$^2$ incident pump pulse).

Figs. 4(a) and (b) depict the theoretically calculated transient enhancements in the magnetic asymmetry following laser excitation, compared with experiment (in grey). We see that the enhancements predicted by the TDDFT calculations for Co$_2$MnGa (Fig. 4(a)) at many energies over the Co-edge are longer lived and extend over a broader energy range than for Co$_2$MnGe (Fig. 4(b)). For example, in the region of 58-60 eV, the Co$_2$MnGa shows a theoretically calculated window of no enhancements with a width of 0.6 eV (Fig. 4(a)) while the window for Co$_2$MnGe is much larger at 1.4 eV (Fig. 4(b)). The additional valence electron in the Co$_2$MnGe system (relative to Co$_2$MnGa) shifts the Fermi energy ≈ 0.4 eV higher, see Fig. 1(a) and (b). This influences which unoccupied final states are available to the pump and the probe, and therefore it changes both the static and dynamic asymmetry. Furthermore, a higher Fermi energy in $Co_2MnGe$ means that the available states in the minority band are closer to the Fermi energy where spin-orbit coupling effects are stronger(*24, 33*) which can also act to reduce spin lifetimes.

Figures 4(c) and (d) show the calculated changes in magnetic moment, for the Ga- and the Ge-based system, respectively. We note that the OISTR induced Co moment enhancement is approximately 2-3 times bigger for $Co_2MnGa$ (Fig. 4(c)) compared to $Co_2MnGe$ (Fig. 4(d)), and is much more long-lived. This can partially explain why experimentally measured enhancements in $Co_2MnGe$ were weaker than in $Co_2MnGa$ and were present at fewer energies. However, the enhancements predicted by TDDFT extend over a larger energy range for $Co_2MnGe$ (Fig. 4) compared to the experimental measurements (gray bars). The discrepancy may have occurred because the theoretical calculation was for a unit cell of the $L2_1$ phase, while the measured sample was in the B2 phase. The TDDFT calculations could only simulate one-unit cell and therefore the effects of crystalline disorder were not taken into consideration in the calculations. The $Co_2MnGe$ sample used in the experiment is also more disordered than the $Co_2MnGa$ system– thus, spin-transfer effects may be weaker in the experiments than what is predicted for the ordered $L2_1$ phase from TDDFT. However, the measured static asymmetries between the B2 and A2 phases were very similar - despite much more significant structural differences between the two (see Supplementary Materials and Fig. S10 for more details).

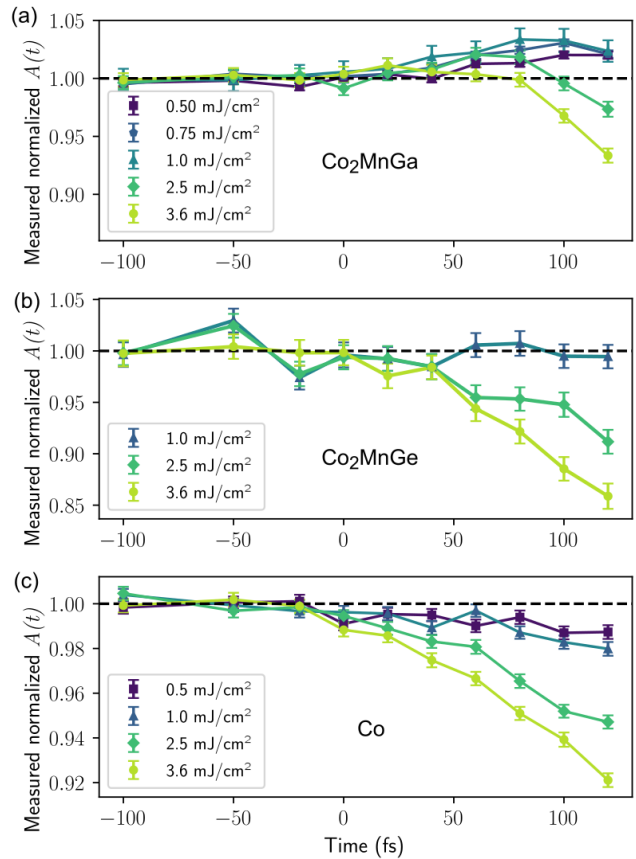

**Figure 5:** Fluence-dependence at the peak of the Co signal for the three samples. The fluences given in the legend correspond to absorbed fluence values. (a) The Co peak (60.4 eV) shows a transient enhancement in $Co_2MnGa$. This enhancement is quenched at higher fluences. (b) At the Co peak in the B2 phase of $Co_2MnGe$ (60.3 eV), there are no perceptible dynamics with 1.0 mJ/cm² pump fluence. No enhancements are seen at any fluence. As the fluence is increased, the demagnetization increases. (c) In the pure Co sample, the demagnetization increases with increasing fluence but no enhancements are seen at the peak asymmetry peak (60.3 eV) at any fluences.

Fluence-dependent measurements for Co$_2$MnGa plotted in Fig. 5(a) show that spin-transfer based enhancements at the Co-peak of Co$_2$MnGa saturate at around 1.0 mJ/cm$^2$, as discussed in Ryan *et al.*(*24*). To verify that we were not in the wrong fluence regime for detecting enhancements of Co$_2$MnGe at the Co-peak, we measured at several different fluences as shown in Fig. 5(b). However, we did not observe enhancements for any of the fluences tested. For each sample, fluence-dependent data was taken with the harmonic comb tuned on resonance with the Co peak (60.4 eV for Co$_2$MnGa and 60.3 eV for both Co$_2$MnGe and pure Co). The fluence-dependent results at other probing energy of Co$_2$MnGe are given in the SM (Fig. S5). Fluence-dependent results at other probing energies of Co had an insufficient signal-to-noise ratio and are not included. Fluence-dependent results at other probing energies of Co$_2$MnGa appear in the supplementary materials of our previous work(*24*).

## DISCUSSION

An important feature of static magnetic asymmetry is the zero-crossing region. The asymmetry is defined as the normalized difference between $I_+$ and $I_-$ reflectivity signals, Eq. 1 in the *Methods* section. Therefore, zero-crossings occur when $I_+$ and $I_-$ are equal. Since the zero-crossings are regions of zero magnetic contrast, small fluctuations around the zero-crossing can lead to unreal enhancements.

We note that in all three samples, we see specific behaviors occurring around the zero-crossing below the Co peak (i.e. in the 55-60 eV range, marked with green dotted vertical lines in Fig. 2). For each of the three samples, we see a region of transient MOKE enhancement just above the zero-crossing, and a region of transient decrease just below the zero-crossing. In the case of Co$_2$MnGa, this manifests as an isolated region of no enhancement just below the zero-crossing, as shown in Fig. 2(a). For the Co$_2$MnGe and pure Co samples, this leads to isolated regions of enhancements above the zero-crossings in Figs. 2(b) and (c). In the pure Co sample, this is the only region of enhancement as depicted in Fig. 3(c). In all three samples, the enhancements in this energy regime can be explained by a red-shifting of the zero-crossing. Several previous works have identified red-shifting of the magneto-optical signal following laser excitation and have explained the possible mechanisms(*8, 27, 34–37*). In the vicinity of the zero-crossing, the asymmetry curve is steep and the absolute magnitude of the MOKE signal is small. If the zero-crossing shifts to lower energies (i.e. redshifts), then the absolute magnitude of positive valued asymmetries will be increased, and the absolute magnitude of negative valued asymmetries will be reduced. This can lead to extremely large measured dynamic enhancements within this energy window due to normalization to nearly zero values, which do not correspond to a real spin transfer signal. As an example of such an artifact, an enhancement of 1000% which was measured for Co$_2$MnGe at 58.5 eV, is given in the supplementary materials, Fig. S4.

Previous studies have shown transient enhancements measured in pure Ni in the regime just above the zero-crossing(*19, 25, 27*). Similarly, multiple OISTR studies have identified a transient enhancement of Ni in an FeNi alloy but only at energies just above the zero-crossing(*23, 25*). Furthermore, one such study(*25*) found that when the angle of incidence of the probe was changed this shifted the zero-crossing energy from below the probe energy to above the probe energy. At angles where the zero-crossing was above the probe energy, enhancements were replaced with very fast reductions in asymmetry. This agrees with the measurements in this study. For example, for Co$_2$MnGe, we see a very rapid decrease in the MOKE signal at 58.3 eV (below zero-crossing) and a very rapid enhancement in the MOKE signal at 58.5 eV (above zero-crossing), see panel 1 of SM Fig. S4.

No enhancements were seen outside of the zero-crossing regime for the pure Co sample. We postulate that iso-element spin-transfer-based MOKE enhancements may not be measurable in conductors with the pump and probe durations used in this study. Indeed, enhancements predicted by TDDFT for the pure Co sample were very small despite many available excitations in the spin-down channel. In metallic samples, excited electrons can rapidly scatter to lower energy states. However, in half-metallic samples such as $Co_2MnGa$ and $Co_2MnGe$, the excited minority spin lifetimes are extended by the insulating gap. Minority spins excited above a half-metallic gap may then be "trapped" by the insulating gap, and the extended lifetimes then allow for a measurable spin-transfer signal to build up throughout the laser excitation pulse.

In fact, throughout the literature, the half-metallic Heuslers $Co_2MnGa$ and $Co_2MnGe$(*22*, *24*) appear to be the only compounds in which spin-transfer based MOKE enhancements have been measured away from the zero-crossing with element-resolved techniques. Transient MOKE enhancements with visible probes (non-element specific) have also been observed for half-metallic half-Heuslers(*36*). This leads us to the conclusion that half-metals may be uniquely capable of sustaining excited OISTR electrons long enough to induce a measurable enhancement in the MOKE signal due to spin transfer.

In electronic structure calculations of the $L2_1$ ordered phase of $Co_2MnGa$ and $Co_2MnGe$, the well-defined atomic arrangement fosters coherent electronic states, where electron-electron interaction is the dominant scattering mechanism. Due to the distinct half-metallic gap and coherent electronic states, we estimate the shortest lifetime of the conduction electrons near the Fermi level for Co and Mn in the fully ordered $L2_1$ phase to persist for approximately 300 fs (Fig. S10). This estimate was made using dynamical mean field theory, as detailed in the supplementary material. The calculations of the $L2_1$ ordered phase highlight the critical role of the half-metallic gap in prolonging the lifetimes of excited electrons, suggesting that the gap in half-metallic systems such as well-ordered $Co_2MnGa$ effectively "traps" minority spins excited above the gap. Thus, for the $L2_1$ ordered phase, the estimated lifetimes (>300 fs) are comparable or greater than the time frame of the observed enhancement shown in Fig. 3(a) for $Co_2MnGa$. Hence, these long-excited electron lifetimes make it possible to observe a strong spin-transfer signal in this material.

In addition, the increased disorder of the B2 and A2 phases (Fig. S8), compared to the $L2_1$ phase, induce a high density of impurities and defects which act as scattering centers. This results in substantial modifications in the electronic structure which produces less coherent electronic states, leading to significantly shorter electron lifetimes in $Co_2MnGe$. We have investigated this by means of alloy theory, as implemented in the coherent potential approximation (CPA), and the calculated electronic structures, as well as life-times, are given in the supplementary materials (Figs. S9 and S10). The disorder inherent in the $B_2$ phase of $Co_2MnGe$ globally smooths out the features present in the $L2_1$ phase band structure with a minor narrowing of the minority bandgap compared to the $L2_1$ phase(*38*). The presence of localized defects can trap electrons, hindering their mobility and leading to a less coherent electronic response. Our calculations estimate electron lifetimes in the $B_2$ phase at 40 fs (more details are given in the Supplementary Materials). In the fully disordered $A_2$ phase (which did not show any spin-transfer based enhancements in the single energy measurements by Tengdin *et al.*(*22*)), electron scattering pathways dominate, with estimated lifetimes dropping to 1 fs - 2 fs (Fig. S10). Therefore, the role of electron-impurity scattering in disordered samples is in clear contrast with the behavior observed in ordered materials. The estimated lifetimes for the disordered samples are significantly shorter than the timescales measured experimentally which indicates the importance of including disorder effects in the analysis.

We have demonstrated that engineering the band structure of the valence electrons in Heusler compounds is possible through the substitution of non-magnetic elements — in this case, by replacing Ga with Ge. This substitution results in dramatic differences in the energy-dependent pump-induced transient magnetic asymmetry. Previous studies have compared the dynamic responses between different alloys or phases of a magnetic material at one specific probe energy. However, this approach cannot provide a full picture of the relevant physics. Changes in a material's structure or composition will change the DOS, and therefore, impact which specific states are accessible at a given probe energy as well as the lifetimes of these states. For example, despite the similarities between $Co_2MnGe$ and $Co_2MnGa$, the measured location of the zero-crossing moved by almost 0.7 eV between the two samples (see Fig. 2). Similarly, the location of the Co peak moved significantly between the A2 and B2 phases of $Co_2MnGe$ SM (Fig. S11). We also demonstrate how the disorder of a B2 crystal structure in $Co_2MnGe$ reduces the lifetimes of the excited electron states. This is shown in Fig. 3, where the asymmetry of the Ge-system decays quicker than that of the Ga-system. Also, in Fig. 4 the energy interval for observed enhancements of magnetic asymmetry is shorter for the Ge-system. Therefore, one must be cautious when directly comparing two different samples at a fixed probing energy. Moreover, the overlapping region between Co and Mn *M*-edges (from 53 eV to 57 eV) is not suitable for element-specific dynamics. We could also identify the zero-crossing regions in the magnetic asymmetry that are sensitive to small fluctuations e.g. a shift of the magnetic asymmetry or fluctuations in the non-magnetic contributions to the TMOKE signal, see Eq. 1. This was observed for the three materials in this study as well as being replicated in many other ultrafast studies. This signature may be misinterpreted as a spin-transfer effect.

## CONCLUSION

In conclusion, by taking comprehensive energy-tuned EUV TMOKE data on three samples, $Co_2MnGa$, $Co_2MnGe$, and pure Co, and comparing the measurements with TDDFT calculations, we make several new findings about ultrafast spin transfer. First, MOKE enhancements in $Co_2MnGe$ were less pronounced than in those in $Co_2MnGa$, which we attribute to the influence of the partial crystalline disorder of the B2 phase and the concomitant reduced lifetime of the electron states. Theory based on CPA shows that the disorder significantly reduces the lifetimes of excited electrons compared to the fully ordered $L2_1$ phase of $Co_2MnGa$, where the half-metallic gap supports prolonged electron lifetimes. We hypothesize that excited spin lifetimes in less ordered samples are too short for spin-transfer-based enhancements to be measurable for a tens of femtoseconds duration of the pump and probe pulses, making the $L2_1$ phase more suitable for spintronic applications. Second, while intersite spin-transfer effects are not possible in pure Co, strong intrasite spin-transfers were expected in the minority channel. However, enhancements in pure Co were not observed outside of the zero-crossing regime, again due to very short spin lifetimes in a metal. Third, probing the ultrafast spin dynamics at one specific energy in an alloy or compound does not provide conclusive information about the ultrafast spin dynamics induced by an ultrafast pump laser. Instead, a full energy scan for the dynamics at the *M*-edge is crucial in the comparison of the ultrafast dynamics of similar alloys due to the modulations of the valence band structure. Finally, the zero-crossing regions of the magnetic asymmetry are sensitive to small fluctuations e.g. a shift of the magnetic asymmetry or fluctuations in the non-magnetic contributions to the TMOKE signal. This signature could be misinterpreted as a spin-transfer effect and thus these regions should be avoided.

## METHODS

Measurements were made at the X-MATTER beamline(*39*). We utilize high harmonic generation (HHG) to generate extreme-ultraviolet light in the energy regime of the *M*-edges of Mn and Co. Measurements were made using the transverse-magneto optical Kerr effect (TMOKE) at a 50° angle of incidence (measured from normal-incidence). 775 nm - 805 nm pulses with a duration of 40 fs - 55 fs generated from a Ti: Sapphire laser were used to excite the sample dynamics (pump) as well as to generate the HHG beam (probe). More details of the experimental design and methodology are outlined in our previous studies(*24*, *39*).

Four samples are utilized in this study: (a) a 20 nm $Co_2MnGa$ sample grown in the $L2_1$ with a 3 nm Si capping layer, (b) a 10 nm $Co_2MnGe$ samples grown in the B2 phase with a 2.8 nm Ta capping layer (oxidized) and (c) a polycrystalline 15 nm Co sample with a 5 nm $Si_3N_4$ capping layer. A description of the $Co_2MnGa$ growth methods is given in Ryan *et al.*(*24*). A description of the $Co_2MnGe$ growth methods is given in Shaw *et al.*(*38*). The TDDFT could only simulate one unit cell and therefore the effects of crystalline disorder were not taken into consideration in the calculations. Therefore, both the $Co_2MnGa$ and $Co_2MnGe$ were simulated for the $L2_1$ phase and the Co sample was simulated as a fully ordered hcp crystal. The specific details of the implementation of TDDFT utilized are described in Ryan *et al.*(*24*).

We note that the sample used for the $Co_2MnGe$ measurements was part of the same sample series as the one used by Tengdin *et al.*(*22*) and are expected to have similar characteristics(*38*). Furthermore, data from the $Co_2MnGe$ sample without any energy shifting of the driving laser(*39*) was in good agreement with the results of Tengdin *et al.* (22).

## TDDFT CALCULATIONS

The TDDFT Calculations were performed for one unit cell of $Co_2MnGa$, $Co_2MnGe$, and Co respectively as implemented in the ELK code(*40*). The calculations were fully ab initio and non-collinear with 10 × 10 × 10 k-points in the Brillouin zone. A time step of 2.4 attoseconds and a smearing width of 0.027 eV was used for the ground state and for time propagation, but for the response function calculations, a Lorentzian smearing width of 0.8 eV was used for the Heusler compounds and a 2 eV smearing was used for the Co. A laser pulse with a wavelength of 800 nm, and full width at half maximum (FWHM) of 45 fs was allowed to interact with each electronic subsystem and the responses to this external field were followed for over 90 fs. More details of the theoretical framework are available here(*24*).

## CALCULATING MAGNETIC ASYMMETRY

The TMOKE experiment probes the magnetic state of the sample through the measurement of the magnetic asymmetry, A. The magnetic asymmetry is defined from the intensity of the reflected EUV beam with the sample magnetized in two opposite directions, denoted '+' and '-'(*14*, *41*):

$$A = \frac{I_+ - I_-}{I_+ + I_-} \quad (1)$$

where $I_+$ is the reflected intensity for the positive magnetization direction, and $I_-$ is the reflected intensity for the negative magnetization direction. The reflected intensities are related to the magnetic (off-diagonal) contribution to the dielectric tensor, $\epsilon_{xy}$, by the following relationship:

$$I_\pm = |I_0|^2 + I_m \epsilon_{xy} \pm \text{Re}\{2I_0 I_m \epsilon_{xy}\} \quad (2)$$

where

$$I_0 = \frac{n \cos \theta_i - \cos \theta_r}{n \cos \theta_i + \cos \theta_r} \quad (3)$$

and

$$I_m = \frac{\sin 2\theta_i}{n^2(n\cos\theta_i + \cos\theta_r)}. \tag{4}$$

$\theta_i$ is the incident angle of incidence, $\theta_r$ is the incident angle of refraction, and $n$ is the refractive index of the material.

## COHERENT POTENTIAL APPROXIMATION (CPA)

Chemical disorder in the material introduces variations in the local atomic environment, leading to scattering of quasiparticles. This scattering disrupts the coherence of quasiparticle states, shortening their lifetimes and broadening their energy levels. In this work, we use the Coherent Potential Approximation (CPA) *(40, 41)* implemented in the SPRKKR package *(42)* to model the effects of chemical disorder by replacing the disordered lattice with an effective medium whose scattering properties represent the configurational average of the different atomic species. CPA provides a self-consistent solution to the effective medium's self-energy, which includes the contributions from all (local) scattering processes due to disorder. The imaginary part of the self-energy provided by CPA directly yields the disorder-induced lifetime broadening, allowing for a quantitative description of how chemical disorder affects quasiparticle lifetimes. For details see the SM section.

## DMFT CALCULATIONS

We have carried out a fully relativistic DFT+DMFT investigation, allowing for an explicit treatment of electron-electron interactions, and direct access to the spectral function and self-energy that captures the quasi-particle life-time. In the presence of a significant electron-electron correlation, the total Hamiltonian of a system is written in terms of an effective Hubbard-model that is mapped in DMFT to a multiband Anderson impurity model that is solved by using the spin-polarized T-matrix fluctuation-exchange (SPTF) impurity solver. For details see the SM section.

## AKNOWLEDGMENTS

**Funding:** The experimental measurements were supported by the Department of Energy Office of Basic Energy Sciences X-Ray Scattering Program under award no. DE-SC0002002. The computations were enabled by resources provided by the National Academic Infrastructure for Supercomputing in Sweden (NAISS), partially funded by the Swedish Research Council through grant agreement no. 2022-06725.

**Competing interests:** H.C.K. and M.M.M. have a financial interest in a laser company, KMLabs, that produces the lasers used in this work. H.C.K. is partially employed by KMLabs. All other authors declare that they have no competing interests.

Certain equipment, instruments, software, or materials are identified in this paper in order to specify the experimental procedure adequately. Such identification is not intended to imply recommendation or endorsement of any product or service by NIST, nor is it intended to imply that the materials or equipment identified are necessarily the best available for the purpose.

# SUPPLEMENTARY MATERIALS: INFLUENCE OF ELECTRON SCATTERING AND SPIN-ORBIT COUPLING ON LIGHT-INDUCED SPIN TRANSFER

In this supplementary material, we provide additional experimental datasets obtained at various probing energies and fluences, expanding upon the data presented in the main manuscript. We also include computational results from time-dependent density functional theory (TDDFT) that explore changes in electronic occupation and magnetic asymmetry over time. Furthermore, we present spectral function calculations from dynamical mean field theory (DMFT) to estimate the lifetimes of excited electrons in the $L2_1$ phase. Data derived from the Coherent Potential Approximation (CPA) for a more disordered B2 phase and the fully disordered A2 phase are also provided in addition to a comparison of electron lifetimes in the B2 and A2 phases for $Co_2MnGa$ and $Co_2MnGe$.

Figures S1 – S4 demonstrate the wide variety of demagnetization rates, as well as the strength and duration of MOKE enhancements, measured across a wide range of probing energies for the pure Co and $Co_2MnGe$ samples. Similar data for $Co_2MnGa$ appears in the Supplementary Materials of our previous study[1]. These figures show that probing at one EUV energy is not sufficient in determining a complete picture of a material's ultrafast magnetic response. Notably, enhancements are seen in the pure Co sample only at the zero-crossing but enhancements appear at additional energies in the compounds ($Co_2MnGe$ and $Co_2MnGa$).

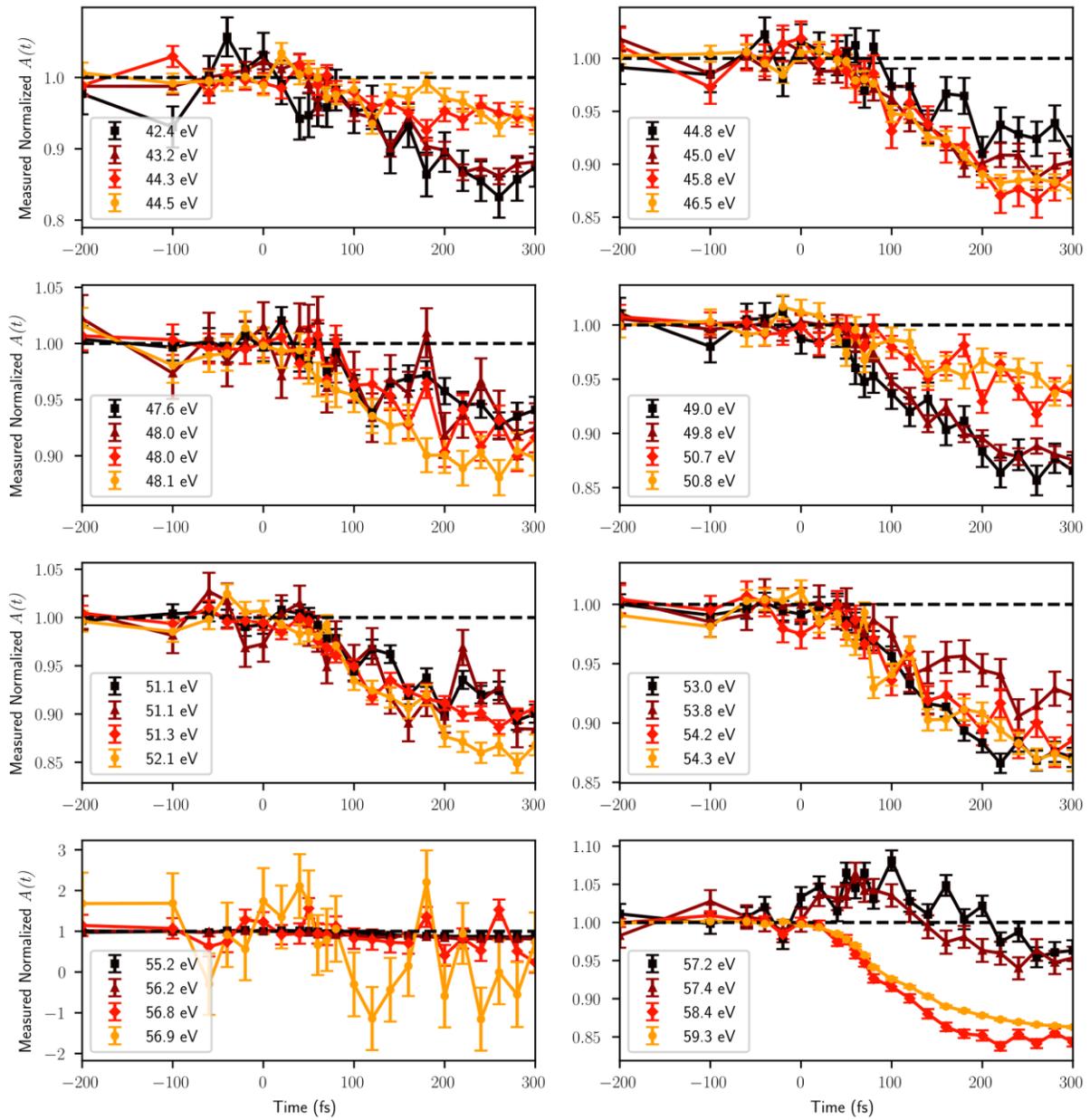

**Figure S1:** Energy-dependent Co dynamics from to 42.4 eV to 59.3 eV. The incident laser fluence was 3.4 mJ/cm$^2$ corresponding to an absorbed fluence of 2.5 mJ/cm$^2$.

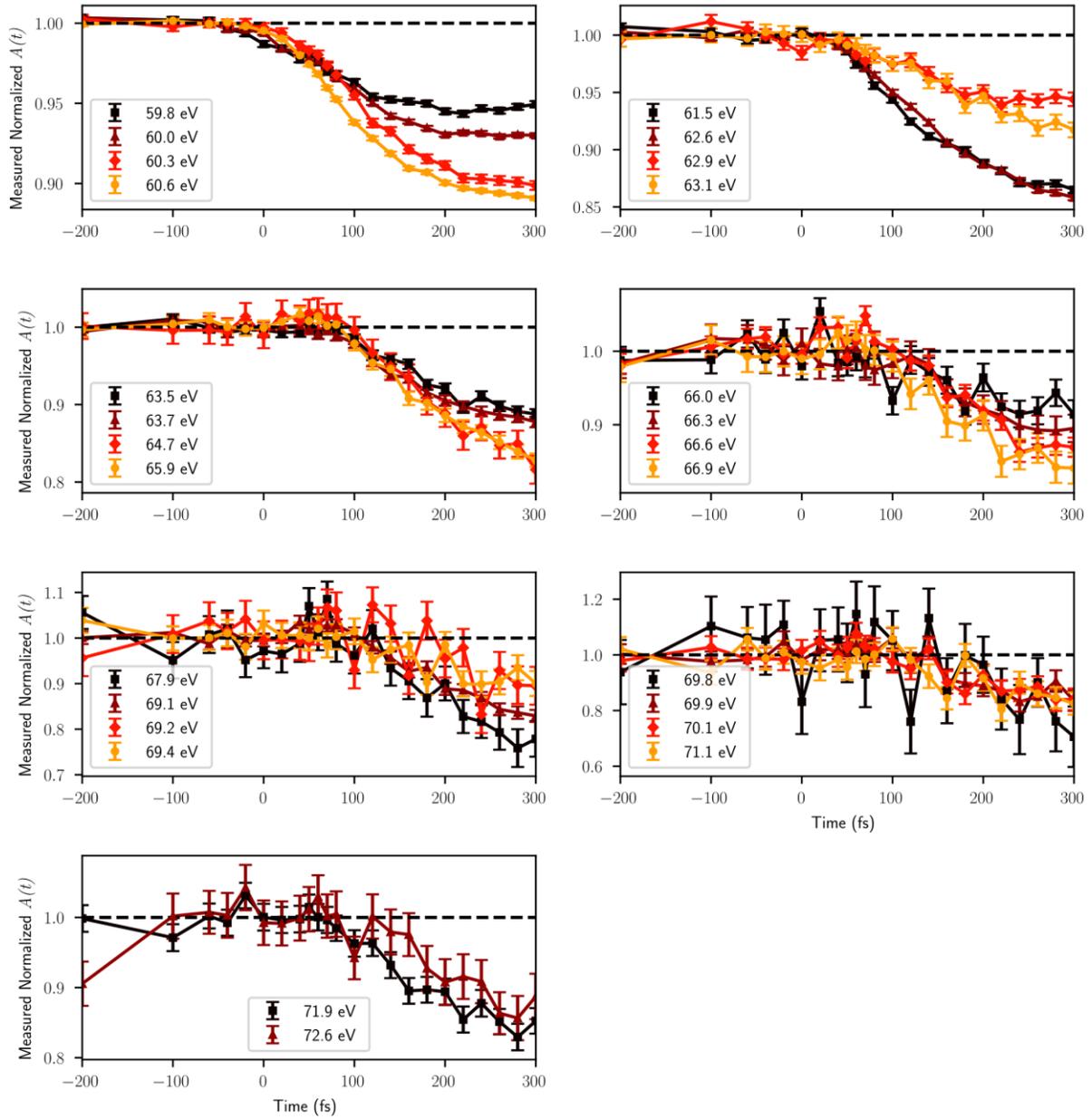

**Figure S2:** Energy-dependent Co dynamics from to 59.8 eV to 72.6 eV. The incident laser fluence was 3.4 mJ/cm$^2$ corresponding to an absorbed fluence of 2.5 mJ/cm$^2$.

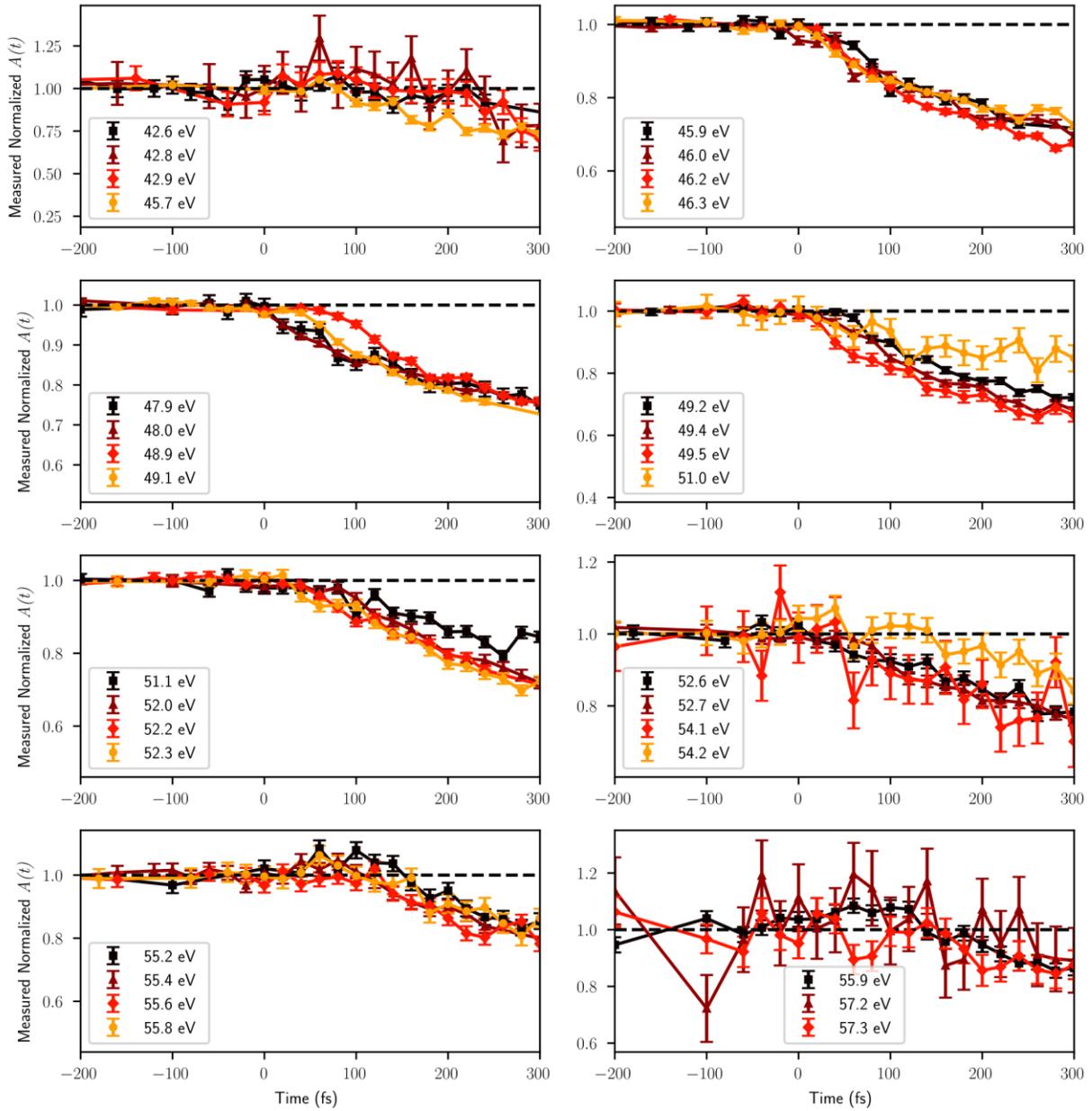

**Figure S3:** Energy-dependent Co$_2$MnGe dynamics from to 42.6 eV to 57.3 eV. The incident laser fluence was 3.4 mJ/cm$^2$ corresponding to an absorbed fluences of 2.5 mJ/cm$^2$.

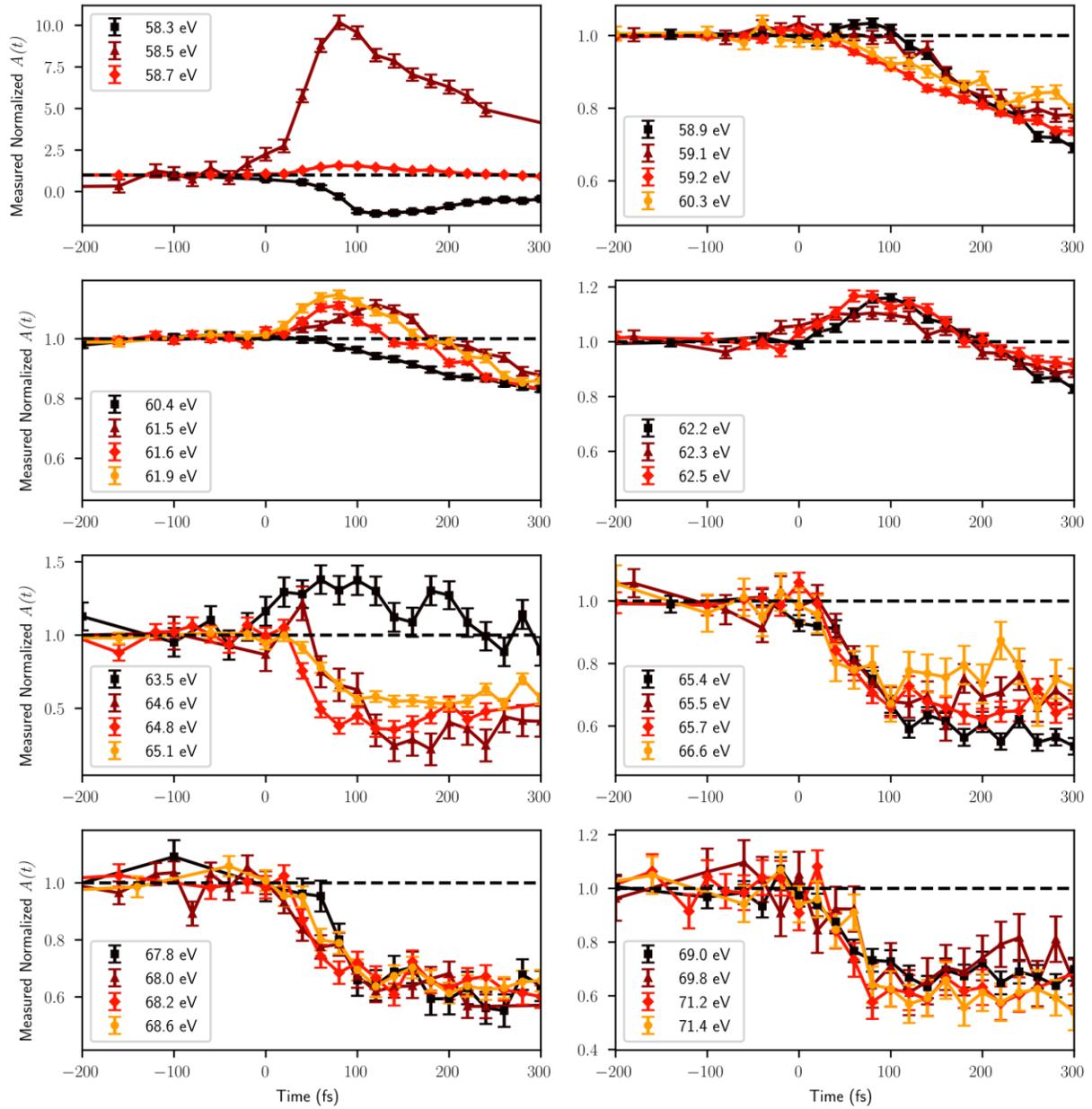

**Figure S4:** Energy-dependent Co$_2$MnGe dynamics from to 58.3 eV to 71.4 eV. The incident laser fluence was 3.4 mJ/cm$^2$ corresponding to an absorbed fluences of 2.5 mJ/cm$^2$.

We measured the changes in the asymmetry of $Co_2MnGe$ at several fluences to determine whether a MOKE enhancement at the Co peak could be measured at any fluence, as depicted in Fig. 5 of the main text. Fig. S5 includes several other energies probed by the EUV harmonics obtained simultaneously during the fluence-dependent measurements of $Co_2MnGe$ as depicted in Fig. 5. While $Co_2MnGe$ exhibits less demagnetization than Co at 1.0 mJ/cm$^2$, it exhibits a much larger percentage of demagnetization than Co at a fluence of 3.6 mJ/cm$^2$. This is evidence that there could be competing effects in the $Co_2MnGe$ signal (spin-transfer based MOKE enhancements vs. demagnetization-based MOKE reductions) which gives a more complex fluence-dependence compared to the pure Co sample. In the pure Co sample, the percentage demagnetization follows a more linear trend with increasing fluence. $Co_2MnGe$ has a Curie temperature, $T_c$, of 905 K(*2*) which is higher than that of $Co_2MnGa$ at 694 K(*2*). However, neither of these Heusler systems have a $T_c$ as high as bulk Co, 1403 K(*3*)**.**

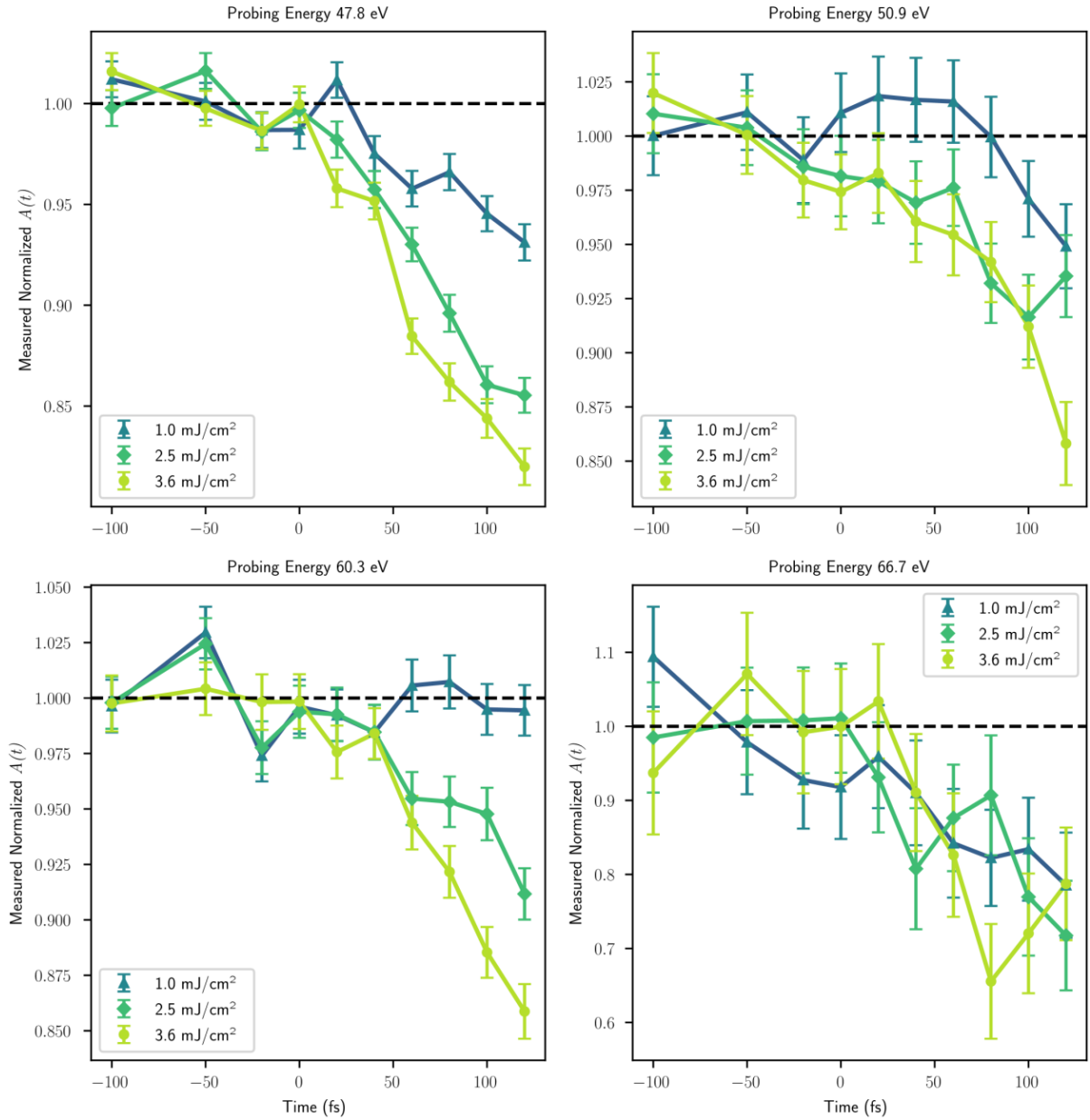

**Figure S5:** Fluence- and energy-dependent $Co_2MnGe$ dynamics. The fluences given in the legend correspond to absorbed fluence values.

Figure S6. provides a more detailed picture of where electrons have been excited to (or depleted from) throughout each material's density of states, 80 fs after the initial onset of the pump pulse. The calculated weighted changes in population (ΔPDOS) for the magnetic sublattices of the three samples: (a) $Co_2MnGa$, (b) $Co_2MnGe$, and (c) pure Co, at 80 fs are given relative to the static case (t = 0). Negative values indicate transient electron depletion at the corresponding energy, while positive values represent transient population increases. The ground-state PDOS for each of the three samples is provided in Fig. 1 of the main text for comparison.

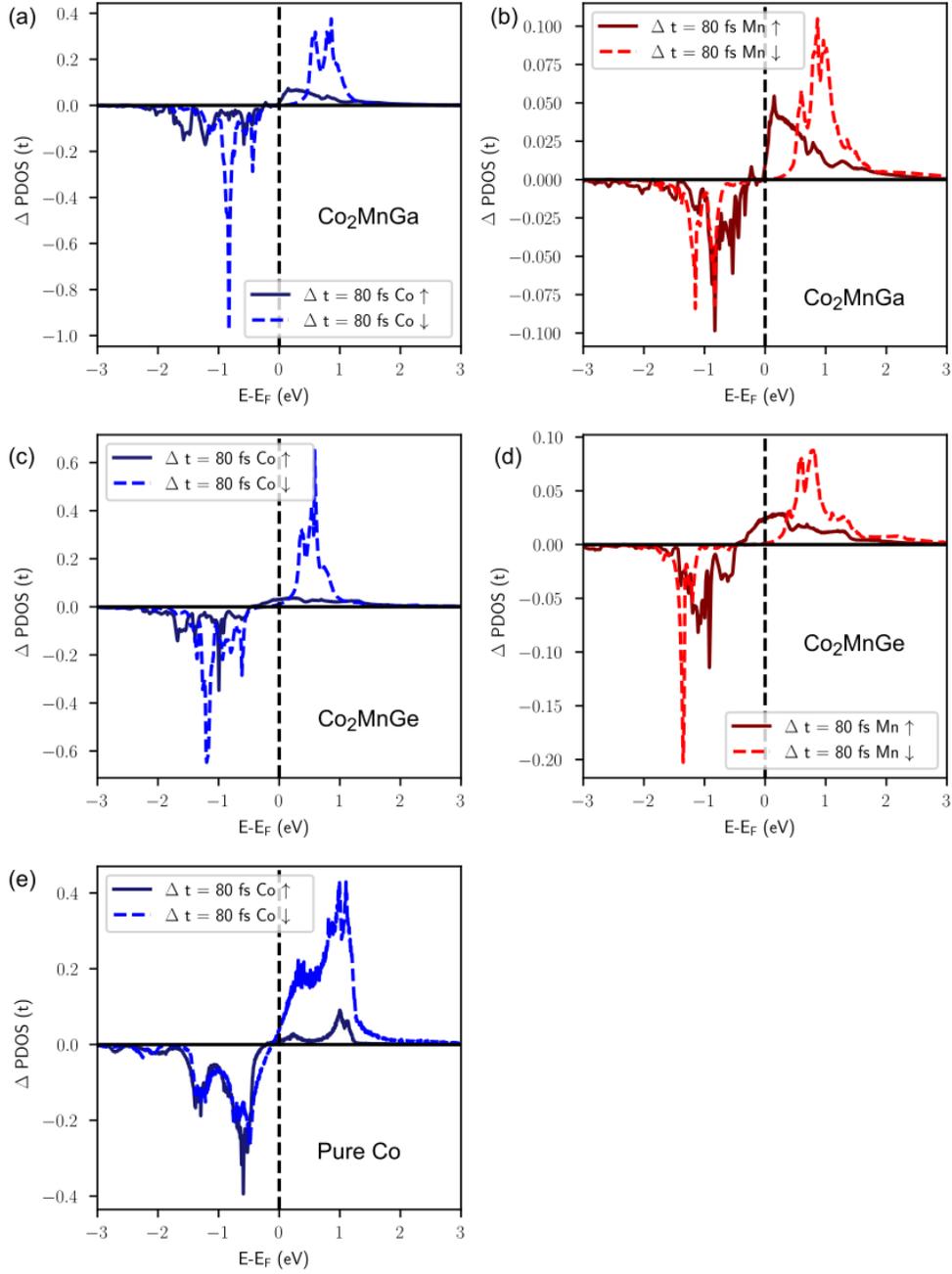

**Figure S6:** The calculated weighted changes in population (ΔPDOS) for the magnetic sublattices of the three samples (a,b) $Co_2MnGa$, (c,d) $Co_2MnGe$ and (e) pure Co. Note that for the compounds, ΔPDOS is shown separately for Co and Mn. Negative values indicate a transient depletion of electrons at the corresponding energy, while positive values indicate a transient population increase at the corresponding energy. The ground state PDOS for each of the three samples are shown in Fig. 1 of the main text. Note the differing y-axis scaling between the Co and Mn subfigures. The majority of changes in population occur in the Co channel for all three samples.

For the two Heusler compounds, Co minority-spin 3d states below the Fermi level exhibit significant changes, reflecting a higher availability of occupied states for pump-induced excitations. These changes predominantly occur within the Co sublattice, as seen in panels (a) and (c) of Fig. S6, and are linked to excitations into Mn minority states, shown in panels (b) and (d). This interspecies excitation pattern suggests a transfer of minority spins from Co to Mn, which results in enhanced Co magnetization and a simultaneous reduction in Mn magnetization, consistent with the simulated magnetic moment changes shown in Fig. 4 of the main text.

In pure Co, panel (e), the minority-spin channel above the Fermi level hosts empty states that are accessible to the pump laser, leading to transient depletion of minority-spin electrons below the Fermi level and a population of states above it. This process is classified as intrasite optical excitation, as discussed in Ryan *et al.*(*1*)

Fig. S7 shows the calculated asymmetry from TD-DFT for the three materials at several different times. Energy-dependent enhancements and decreases in the MOKE signal are depicted across the energy range that was investigated experimentally. In $Co_2MnGe$, no enhancements are seen at the Co peak itself as depicted in Fig. 2 of the main text. Thus, the actual magnitude of the spin transfer observed in the original $Co_2MnGe$ spin transfer study(*4*) is likely less than the 10% value quoted. The measured MOKE enhancement is very different for each probed energy. The absorbed pump fluence used in the $Co_2MnGe$ measurements as well as the TD-DFT calculations was 2.5 mJ/cm$^2$ (which is similar to the value used in the original paper of 2.4 mJ/cm$^2$). The calculated increase in the Co moment from TDDFT is only about 1% at its maximum, as shown in Fig. 4(d) of the main text.

The TD-DFT calculations for the $L2_1$ phases of $Co_2MnGe$ and $Co_2MnGa$ show a significantly smaller and more short-lived magnetic enhancement of the $Co_2MnGe$, which we attribute to the shift in the Fermi energy based on the substitution of Ga with Ge. Furthermore, same-site spin-transfers that were predicted in the pure Co sample showed no enhancement signatures (excluding at the zero-crossing) in this highly disordered metallic sample (where excited electron lifetimes are very short).

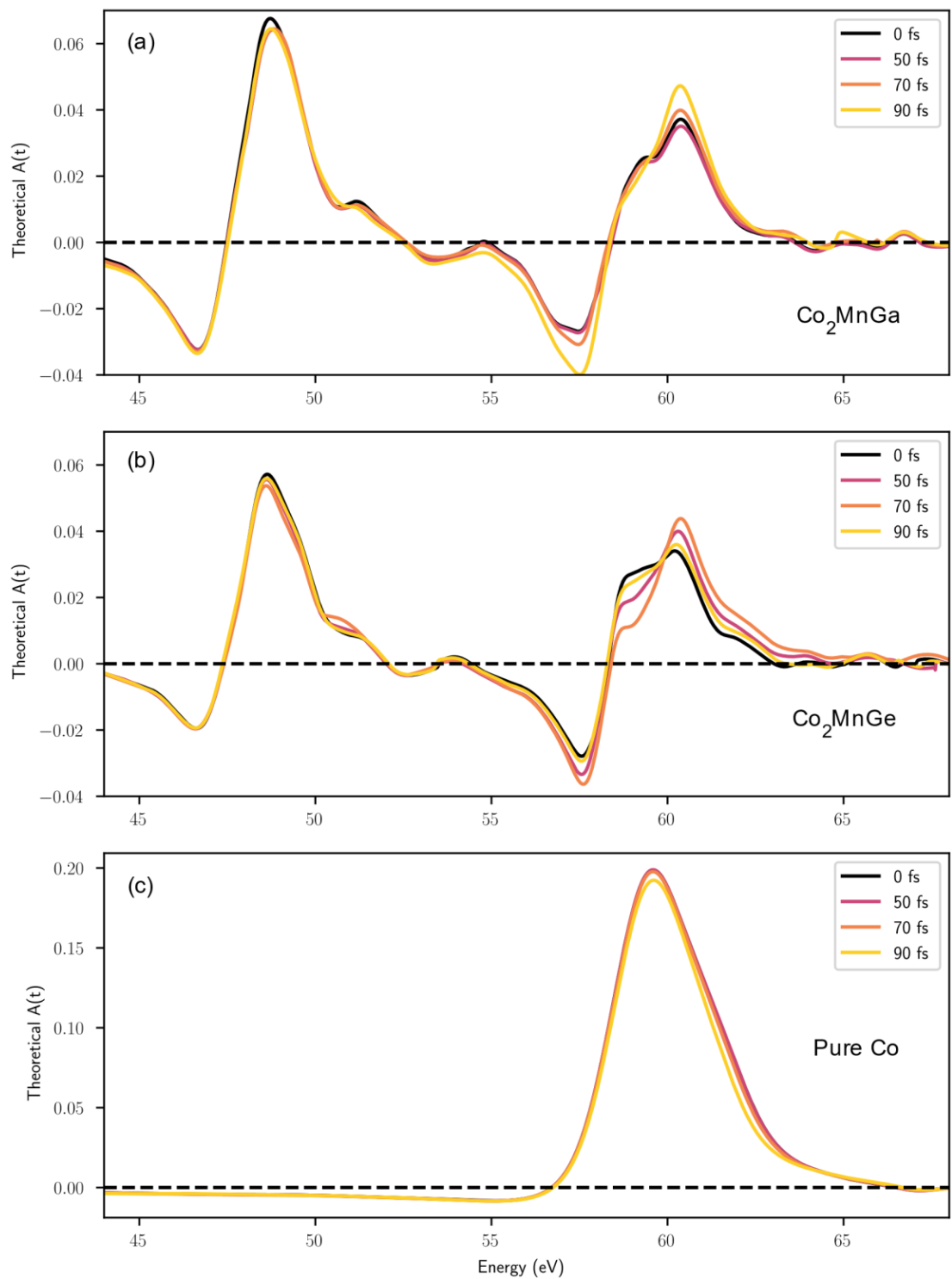

**Figure S7:** Time- and energy-dependent asymmetry curves calculated from TDDFT. The time delays chosen correspond to those used to demonstrate the theoretical enhancements in Figs. 4(a), (b), and (c) of the main text.

There are important differences in the electronic structure between $Co_2MnGe$ and $Co_2MnGa$ beyond the band-filling effects discussed above. In particular, there are lifetime effects induced by electron-electron interaction outside of what is described by density functional theory. Here, we investigate these effects with dynamical mean field theory (DMFT) as well as calculations of the broadening induced by crystalline disorder. As shown in Fig. 8, electron-electron interactions (calculated with DMFT in the $L2_1$ phase) increasingly broaden electron states as their energies move further from the Fermi level. The lifetime broadening at the Fermi level is divergent due to Fermi liquid properties of the DMFT impurity solver (see below details of these calculations). It is important to note that the lifetimes of electron states, as governed by electron-electron interaction, is in excess of the experimental timescale of 300 fs for the ordered $L2_1$ phases of both $Co_2MnGe$ and $Co_2MnGa$. Hence, the dynamical correlation, and the corresponding broadening of electron states, does not drastically influence the magnetization dynamics. Furthermore, the calculated lifetimes of spin-up and spin-down electronic states do not significantly differ (data not shown). We note that the lifetime ($\tau$) was extracted using the imaginary part ($\Im$) of the calculated self-energy $\hat{\Sigma}(R, z)$ using the relation:

$$\tau = \frac{1}{\Im\left(\hat{\Sigma}(R, z)\right)} \tag{S1}$$

More details of the DMFT calculations are given at the end of the supplemental materials.

A key distinction between the Heusler samples lies in the level of crystalline disorder observed in $Co_2MnGe$, which has the B2 phase, compared to the fully ordered $L2_1$ phase of $Co_2MnGa$. While the overall electronic structure and total magnetic moments of the $L2_1$ and B2 phases of $Co_2MnGe$ are nearly identical—differing by less than 0.01% (*5*)—the A2 phase (characterized by full intersite disorder) exhibits a remarkably different electronic structure. In the A2 phase, Co anti-site disorder disrupts the local half-metallicity, eliminating the half-metallic gap and significantly reducing spin polarization at the Fermi level (*5*, *6*). Additionally, the spectral function in this phase displays distinct features, as shown in Fig. S9. It is clear that crystalline disorder strongly impacts the coherence of the electronic states, enhances electron-impurity scattering, and shortens the lifetimes of excited electrons across the structural phases of $Co_2MnGa$ and $Co_2MnGe$. Fig. S9 highlights how the electronic structure evolves with crystalline disorder effects, starting from states with infinite lifetimes in the $L2_1$ phase (based on DFT, see Fig. S9 (a and b)) and progressively becoming broader, first due to disorder between Mn and Ge/Ga (B2 phase) and finally with full disorder in the A2 phase. Notably, the quasi-particle nature is retained in the B2 phase, with a clear connection between energy and crystal momentum, albeit with significant broadening effects. As an example, Fig. S10(a) shows the spectral function at the Γ point for the B2 and A2 phases of $Co_2MnGa$ and $Co_2MnGe$ from -2 to 2 eV relative to the Fermi energy. The peaks in the spectral function of the B2 phase are sharp, while those for the disordered A2 phase are very broad. Note that the disordered A2 phase is metallic. To estimate the difference in the lifetimes of the excited carriers above the Fermi level, we fit the peaks above the Fermi level to a Lorentzian function for both the A2 and B2 phases. Using the Heisenberg uncertainty relation,

$$\tau = \frac{\hbar}{\gamma}, \tag{S2}$$

we convert the width of the fitted Lorentzian ($\gamma$) into an estimate for the lifetime ($\tau$). The width of the sharp Lorentzian peak for the B2 phase is approximately $\gamma \approx 0.016$ eV, corresponding to a lifetime of $\tau \approx$ 41 fs, and fit quality $R^2 \approx 0.87$. In contrast, the broad peak for the A2 phase has a width of $\gamma \approx 4.25$ eV, resulting in a lifetime of $\tau \approx 0.15$ fs. This demonstrates that the lifetime of the more ordered B2 phase is almost two orders of magnitude longer than that of the A2 phase. This is an order of magnitude shorter than the lifetime calculated from electron-electron scattering, as described in the previous section.

In the A2 phase, the intense scattering disrupts the quasi-particle band description. The structural disorder profoundly influences the energy states and, therefore, the dynamics of laser excited electrons. The scattering due to disorder affects the entire energy range and significantly alters the lifetime close to the Fermi level (in contrast to electron-electron scattering). Computational details on the CPA calculations are given at the end of the Supplemental Materials.

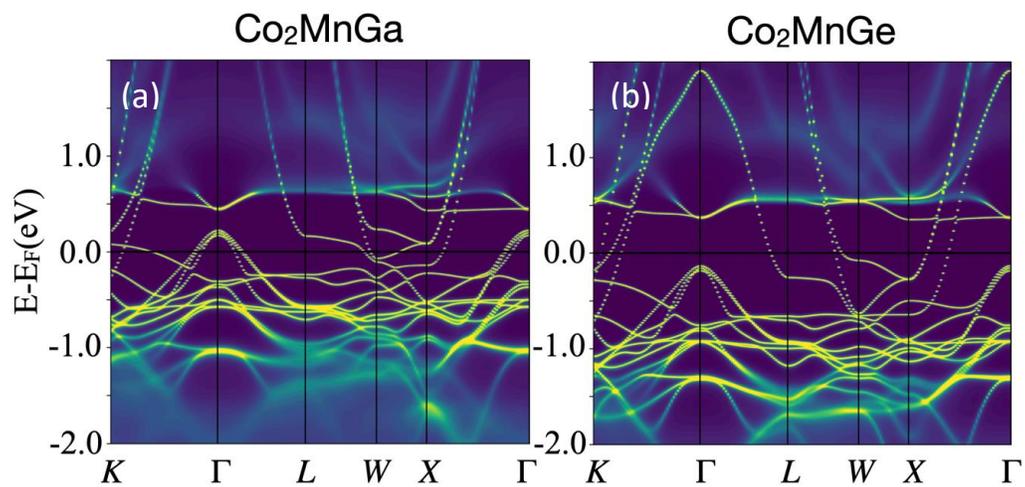

**Figure S8:** DMFT spectral function calculations for (a) $Co_2MnGa$ and (b) $Co_2MnGe$.

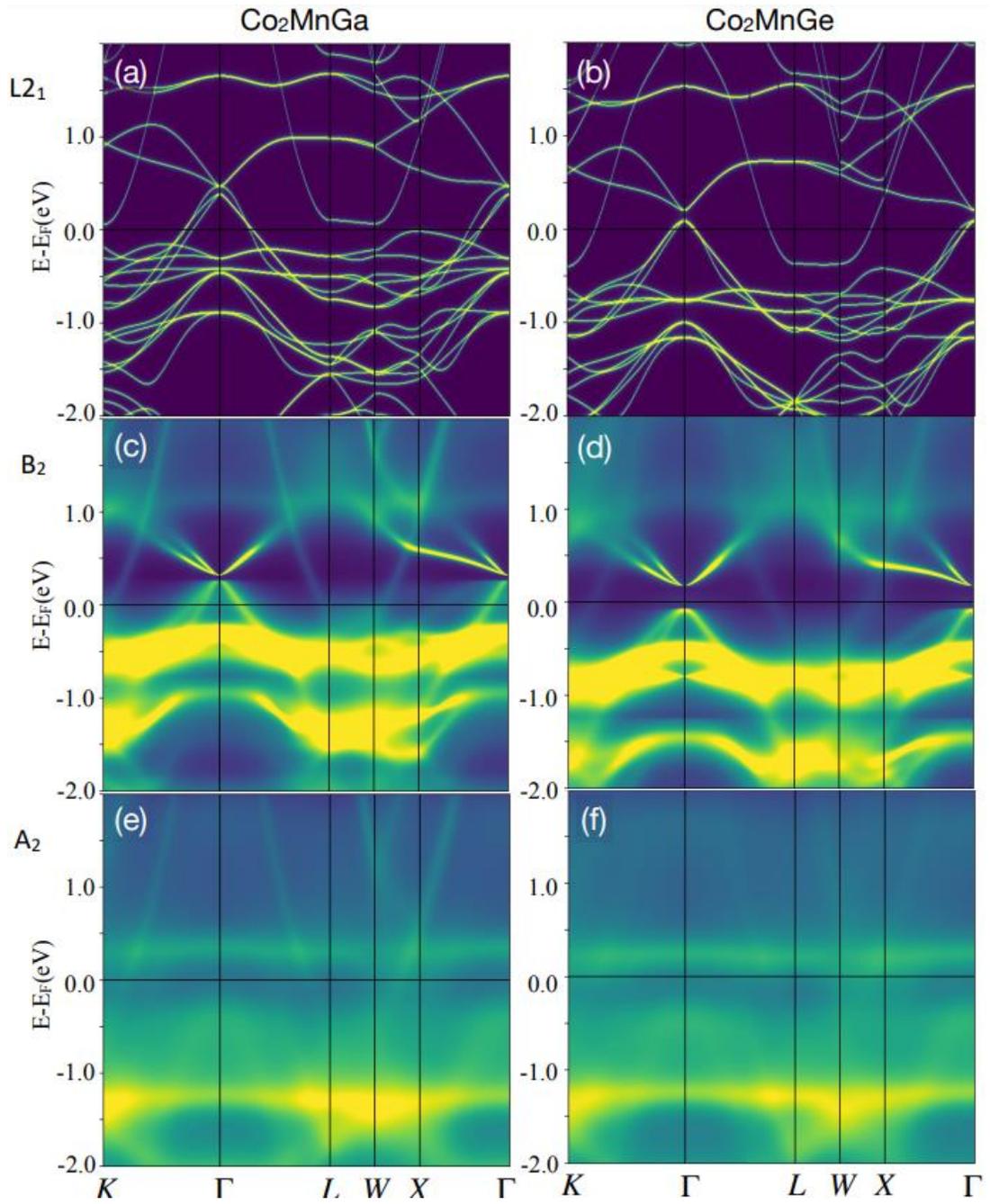

**Figure S9:** Band structures and spectral functions computed using DFT (L2$_1$ phase) and CPA methods (B2 and A2 phases) for Co$_2$MnGa and Co$_2$MnGe. Panels (a), (c), and (e) show the L2$_1$, B2, and A2 phases of Co$_2$MnGa respectively, while panels (b), (d), and (f) present the corresponding data for Co$_2$MnGe. The spectral functions (CPA) reveal the progressive broadening of electronic states as disorder increases from the L2$_1$ to the A2 phase.

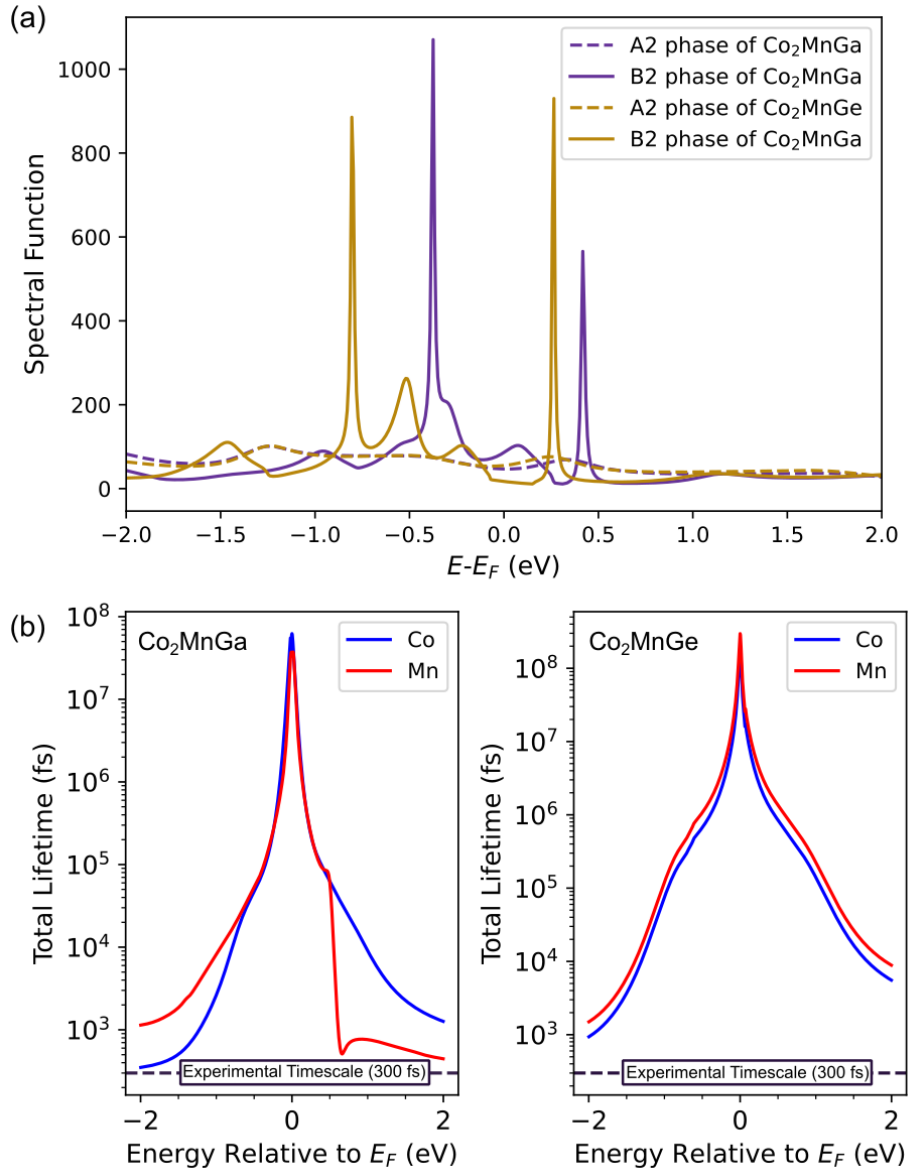

**Figure S10.** (a) Spectral functions for the B2 and A2 phases of Co₂MnGa and Co₂MnGe at the Γ point. (b) Lifetimes were obtained from the DMFT calculations. The experimental measurements in the main text are limited to the first 300 fs following laser excitation. This experimental timescale is represented by purple dotted lines at the bottom of (b) and labelled "Experimental Timescale (300 fs)".

The ground state optical asymmetry of the A2 phase of $Co_2MnGe$ was measured and is presented in Fig. S11. Both the A2 and B2 phases were grown using identical processes, thicknesses, and capping layers allowing a direct comparison of their magnetic properties. The A2 phase sample was more disordered due to a reduced annealing time. The main differences between the measured asymmetries, Fig. S11, of the A2 and B2 phases are the relative heights and positions of the Co and Mn peaks. This difference in relative heights of Co and Mn peaks is expected based on the calculated magnetic moments per atomic site of the A2 and B2 phases. The magnetic moment for Co is approximately 25% higher for the B2 phase compared to the A2 phase, but 60% lower for Mn(*5*). In comparison, the differences in the calculated moments for Mn and Co between the B2 and $L2_1$ phases are less than 2%(*5*).

Tengdin *et al.*(*4*) measured a dynamic MOKE enhancement at the Co-edge of the B2 phase of $Co_2MnGe$ caused by intersite spin-transfer. However, this enhancement was not present when Tengdin *et al.* measured at the same energy in the A2 phase. We this lack of enhancement to differences in the valence band structure. The most significant difference is absence of the band gap in the minority spin channel of the disordered A2 phase, unlike the ordered half-metallic B2 and $L2_1$ phases.

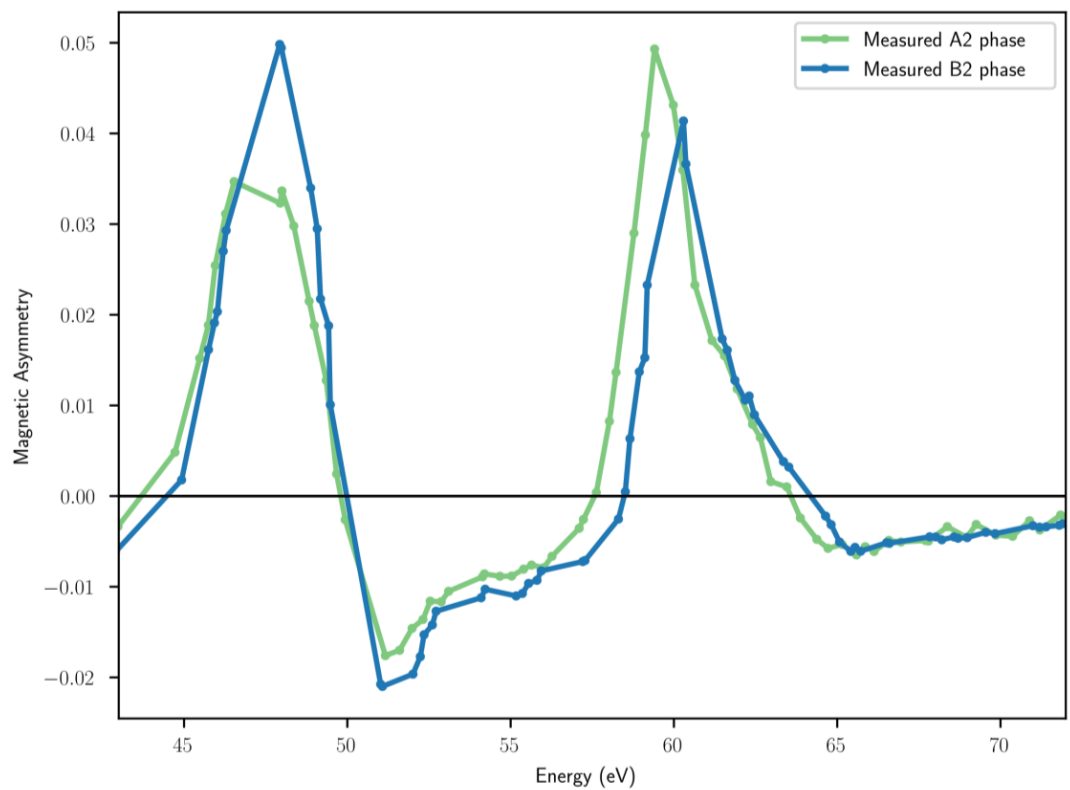

**Figure S11:** Comparison of the measured ground state magnetic asymmetries of the A2 and B2 phases of $Co_2MnGe$. The A2 phase had the same thickness and capping layer as the B2 phase and was grown with the same processes and a reduced annealing time.

## Effects of chemical disorder via the Coherent potential Approximation (CPA)

The effect of the chemical disorder on the band structures for disordered and partially disordered systems was taken into account within the CPA as implemented in the spin-polarized relativistic Korringa-Kohn-Rostoker (SPR-KKR) code(7). The Vosko-Wilk-Nusair(8) version of the local spin density approximation was employed as the exchange-correlation functional. The Bloch spectral functions were derived within the atomic sphere approximation in a scalar relativistic calculation. The irreducible wedge of the Brillouin zone was sampled with 3000 $k$ points and $s$, $p$, $d$, and $f$ orbitals were included in the basis set ($l_{max}$ = 4). The calculations are done for the standardized unit cell of $Co_2MnGe$ and $Co_2MnGa$.

## Effects of electron-electron scattering via Dynamical Mean Field Theory (DMFT)

To capture the impact of strong electron correlation on the quasi-particle lifetimes, we have carried out a fully relativistic DFT+DMFT investigation allowing for an explicit treatment of electron-electron interactions, and direct access to the spectral function and self-energy that captures the contribution to the quasi-particle life-time. In the presence of a significant electron-electron correlation, the total Hamiltonian of a system is written in terms of an effective Hubbard-model as:

$$H = H_{DFT} + \frac{1}{2}\sum_{R}\sum_{\xi_1....\xi_4} U_{\xi_1....\xi_4} c^{\dagger}_{R,\xi_1} c^{\dagger}_{R,\xi_2} c_{R,\xi_3} c_{R,\xi_4} \quad (S3)$$

Here, $H_{DFT}$ is the Hamiltonian computed using DFT with R and $\xi$ being the Bravais lattice site index and the orbital index of the correlated atom respectively. In spectral DFT, the single-electron Green's function is related to the unperturbed DFT Hamiltonian $\hat{h}_{LDA}$ as:

$$\hat{G}(z) = \left[(z-\mu)\mathbb{I} - \hat{h}_{LDA} - \hat{\Sigma}(z)\right]^{-1} \quad (S4)$$

Here z is the energy in the complex plane, μ is the chemical potential and the many-body representative of the electron interaction is written in terms of the self-energy operator, $\hat{\Sigma}(z)$. The main observable of spectral DFT is the local Green's function at the site R:

$$\hat{G}_R(z) = \hat{P}_R \hat{G}(z) \hat{P}_R \quad (S5)$$

Here $\hat{P}_R$ is the projection operator $|R,\xi\rangle\langle R,\xi|$ in the correlated subspace $\{|R,\xi\rangle\}$. Spectral DFT works under Dynamical mean field theory (DMFT) approximation, with a "local" self-energy(9), where at a single "impurity" site R, the effects of the other sites are replaced with a self-consistent electronic bath $G_0^{-1}(R,z)$. Thus, the atomic "impurity" site is embedded in a fermionic bath and that system is treated within the multiband Anderson impurity model. The self-energy operator can be explicitly calculated from the impurity Green's function $\hat{G}_{imp}(z)$ with the help of the inverse Dyson equation after solving the "impurity" problem as:

$$\hat{\Sigma}(R,z) = \mathcal{G}_0^{-1}(R,z) - G_{imp}^{-1}(z) \quad (S6)$$

We have solved this multiband Anderson impurity model by using the spin-polarized T-matrix fluctuation-exchange (SPTF) impurity solver(10, 11). After solving the effective impurity problem, one obtains the updated self-energy and the respective modified chemical potential and thereby a new single-particle

Green's function and a new electronic bath $\mathcal{G}_0^{-1}(R, z)$ by using the inverse Dyson equation:

$$\mathcal{G}_0^{-1}(R, z) = G_R^{-1}(z) + \hat{\Sigma}(R, z) \quad (S7)$$

This full iterative process is continued to obtain a converged self-energy. After embedding this self-energy onto the DFT Hamiltonian and carrying out an electronic charge convergence via the self-consistent cycle, we obtain a complete solution of DFT+DMFT. All of these calculations are done using the DMFT implementation in the relativistic spin polarized toolkit (RSPt)(*12, 13*) which is based on a full-potential description using linearized muffin-tin orbitals. In the present work, the starting point for the DFT+DMFT were DFT calculations with a Hubbard U. In this DFT+U mode, the values of U-parameter for Co and Mn are taken as 3 and 2 eV with the exchange parameter (J) value to be 0.9 for both. The DFT+DMFT calculation was done with the converged potential of the DFT+U step after using the SPTF solver.